\documentclass{article}

\usepackage{titling}
\usepackage[centertags]{amsmath}
\usepackage{chicago,fancyhdr}
\usepackage{amssymb,amsfonts,eucal,oldgerm}
\usepackage[polutonikogreek,french,german,english]{babel}
\usepackage[colorlinks=true,citecolor=red,linkcolor=blue,urlcolor=blue,pdftex]{hyperref}
\usepackage[pdftex]{graphicx}
\usepackage[text={6.1in,8.25in},centering]{geometry}

\newcommand{\skipline}[1][1]{\vspace*{#1\baselineskip}}
\newcommand{\smidge}{\hspace*{.1em}}

\newcommand{\coloneq}{\mathrel{\mathop:}=}

\predate{}
\postdate{\skipline[-1]}

\title{If Metrical Structure Were Not Dynamical, Counterfactuals in
  General Relativity Would Be Easy\thanks{I thank David Malament
    (private correspondence) and Harvey Brown, Dennis Lehmkuhl and
    Oliver Pooley (conversation) for vigorously pressing me on the
    paper's arguments and conclusions.  I thank The Young Guns of the
    Spacetime Church of the Angle Brackets for suffering through a
    previous version of this paper and giving me, as always insightful
    help with a smile.}}

\author{Erik Curiel\thanks{\textbf{Author's address}: Munich Center
    for Mathematical Philosophy, Ludwig-Maximilians-Universit\"at,
    Ludwigstra{\ss}e 31, 80539 M\"unchen, Deutschland; \textbf{email}:
    \href{mailto:erik@strangebeautiful.com}
    {\texttt{erik@strangebeautiful.com}}}}

\begin{document}
\thispagestyle{empty}

\maketitle

\tableofcontents

\skipline

\begin{quote}
  \begin{center}
    \textbf{ABSTRACT}
  \end{center}  
  
  General relativity poses serious problems for counterfactual
  propositions peculiar to it as a physical theory.  Because these
  problems arise solely from the dynamical nature of spacetime
  geometry, they are shared by all schools of thought on how
  counterfactuals should be interpreted and understood.  Given the
  role of counterfactuals in the characterization of, \emph{inter
    alia}, many accounts of scientific laws, theory confirmation and
  causation, general relativity once again presents us with
  idiosyncratic puzzles any attempt to analyze and understand the
  nature of scientific knowledge must face.
\end{quote}

\skipline


\skipline



\section{Pr\`ecis}
\label{sec:precis}

In his elegant, magisterial exposition of the foundations of general
relativity, \citeN[ch.~2, \S1, pp.~120--121]{malament-fnds-gr-ngt}
provides three interpretive principles to endow the mathematical
framework of Lorentzian geometry with physical content:\footnote{I
  follow \citeN{malament-fnds-gr-ngt} in all relevant conventions (the
  signature of the spacetime metric, the definition of the Weyl
  tensor, \emph{etc}.).  The reader should consult that work or
  \citeN{wald-gr} for exposition of all concepts and results about
  general relativity I rely on in this paper, except where explicitly
  noted otherwise.}
\begin{quote}
  For all smooth curves $\gamma: I \rightarrow M$ [where $I \subset
  \mathbb{R}$ is an open interval and $M$ is a candidate spacetime
  manifold]:
  \begin{description}
      \item[(C1)] $\gamma$ is timelike iff{} $\gamma[I]$ could be the
    worldline of a point particle with positive mass;
      \item[(C2)] $\gamma$ can be reparametrized so as to be a null
    geodesic iff{} $\gamma[I]$ could be the trajectory of a light ray;
      \item[(P1)] $\gamma$ can be reparametrized so as to be a
    timelike geodesic iff{} $\gamma[I]$ could be the worldline of a
    \emph{free} point particle with positive mass.
  \end{description}
\end{quote}
(Emphases are Malament's; `C' indicates the proposition pertains to
the interpretation of conformal structure, `P' to projective
structure; (C1) articulates the physical meaning of timelike curves,
(C2) that of null geodesics, and (C2) that of timelike geodesics.)  He
immediately offers four comments and qualifications to address
possible concerns one may have with these propositions as
interpretative principles, touching on questions about the exclusion
of tachyons, the restriction to smooth curves, the status of point
particles in general relativity, and, of most interest for our
purposes, the modal character of the propositions.  He concludes
(\emph{ibid}., p.~122),
\begin{quote}
  Though these four concerns are important and raise interesting
  questions about the role of idealization and modality in the
  formulation of physical theory, they have little to do with
  relativity theory as such.
\end{quote}
I agree with his conclusion in all parts, except for the concern about
the role of modality.  I think there are important problems with
modality in general, and with the understanding of counterfactuals in
particular, peculiar to general relativity as a physical theory,
problems that have gone unremarked in the philosophy and the physics
literature.\footnote{\label{fn:other-thrs}Strictly speaking, as the
  discussion in \S\ref{sec:prob-detail} will make clear, the problem
  is not restricted to general relativity, but rather infects all
  theories with spacetime structure that is dynamical by virtue of a
  particular kind of coupling with matter sources, such as
  $f(R)$-gravity.  (See, \emph{e}.\emph{g}.,
  \citeNP{felice-tsujikawa-fr-theors} for an extended review.)  I
  focus on general relativity as it our most strongly confirmed and
  most widely accepted theory of spacetime structure, and everything I
  say about it can be translated easily into the context of similar
  theories.  The problem does not arise, however, simply by virtue of
  \emph{any} non-trivial relation a theory may posit between geometry
  and matter: the problem does not arise, for instance, in classical
  Yang-Mills theory on Minkowski spacetime.}  Malament's formulation
and discussion of the interpretative principles allows them to be
drawn out with great clarity.\footnote{I emphasize that I am not
  criticizing Malament or trying to draw attention to weaknesses or
  errors in his exposition, quite the contrary.  It is the exemplary
  (and characteristic) clarity, precision and thoroughness of his
  discussion that allows a previously unrecognized problem to be
  brought to light.}

About (C2) he says (\emph{loc}.\@ \emph{cit}.), ``We are considering
what trajectories are available to light rays when no intervening
material media are present---\emph{i}.\emph{e}., when we are dealing
with light rays in vacuo.''  Now, surely we want to talk as well about
the null cones even at those places where matter is present.  In order
to do so, and in order to formulate the analogue of (C2) for those
spacetime regions (in order to give a physical interpretation to the
null cones at those points), we must say something along the following
lines: the null geodesics where matter is present are those paths
light rays would follow if the matter there were removed.  But on its
face, that modal statement makes no sense in the context of general
relativity, because however we make sense of the idea of ``removing
matter'' from a spacetime region, the metric will \emph{eo ipso} be
different in that region from what it was, and it will generically be
the case that the new metric in that region will not agree with the
original metric on what it counts as null vectors, much less on what
it counts as null geodesics, among many other
differences.\footnote{Indeed, in regions of spacetime filled, say,
  with a non-trivial dielectric, there simply are no null curves that
  are ``the possible paths of light rays'' because the full resources
  of electromagnetic theory tell us in this case that, under any
  reasonable construal of ``propagate'', light does not propagate
  through a dielectric with the speed of light in vacuo.

  The same problem arises for timelike curves in regions of spacetime
  already occupied by matter, \emph{i}.\emph{e}., for (C1) and (P1),
  but I focus on the case of null rays to simplify the exposition.}
The distribution of matter in a region of spacetime in large part
informs the metrical structure there, so what sense can be made, in
the context of the theory, in asking what the metrical structure
\emph{would} be if the matter actually there \emph{were not} there?
And now we face the heart of the problem: the lack of a unique vacuum
solution for the Einstein field equation forces an ineliminable
ambiguity in the idea of what it may mean to ``remove matter from a
region of spacetime'', guaranteeing that we have no way to conclude on
any principled basis ``what the metric would then look like there''.

\section{Counterfactuals in Physical Theories}
\label{sec:cfacts-phys-thrs}

One of the simplest and \emph{prima facie} most promising ways to
begin to try to get a grip on the problem, and to look for its
resolution, is to treat the proposed counterfactual changes as
represented by a change in initial conditions, and so to use the
machinery of general relativity's initial-value formulation to fix the
solution.  Indeed, this is the natural and effective way of dealing
with such counterfactuals in all other physical theories.  When one
wants to know how the state of a system would change if some of its
properties or some properties of its environment were to change, one
first writes down the initial-value formulation (or boundary-value
problem, depending on the details of the theory and the problem at
hand), plugs in the original values for all parameters, and calculates
the solution; one then does the same thing using the new,
counterfactually changed values for the parameters, and compares the
new solution to the old.  Easy as pie, and about as philosophically
unproblematic as things can possibly get for a large class of
fundamentally important counterfactuals.

What makes this procedure work in almost all physical theories is the
context of a family of fixed background structures, often
spatiotemporal ones, with respect to which one has natural ways of
identifying and so comparing entities like ``the same quantity of the
same system at the same place and same time, under otherwise different
conditions or in otherwise different states''.  To see how this works
in a simple case, one that has especial relevance to the problem in
general relativity, consider the situation in Newtonian gravitational
theory.  It makes perfect sense in Newtonian theory to reason
counterfactually about the behavior of a given kind of system in the
presence or absence of any other kind of system, since that presence
or absence won't affect the kinematical structure of Newtonian
spacetime: it is always $\mathbb{R}^4$ with a fixed foliation of
simultaneity slices, and a fixed, flat affine structure defining the
possible trajectories of all freely falling bodies
\cite{stein-newt-st}.\footnote{Nothing of importance for the example
  would change if we were to work in the more sophisticated context of
  geometrized Newtonian gravity \cite{malament-fnds-gr-ngt}.  I
  suspect, however, that in generalizations of geometrized Newtonian
  gravity, in which one does not restrict the topology of spacetime to
  $\mathbb{R}^4$ and one does not (effectively) require that the
  simultaneity slices be spatially flat \cite{malament-fnds-gr-ngt},
  then similar problems would arise.  They would, however, arise for
  the same reasons as they do in general relativity: the absence of
  fixed, absolute, background spatiotemporal structure because of
  dynamical coupling with matter (footnote~\ref{fn:other-thrs}).}
There is no problem in principle in computing the counterfactual
change in gravitational forces in a region induced by any
counterfactual changes in the distribution of matter anywhere in the
spacetime.  For example, one may be interested in the question: what
would happen to the orbits of the planets in the Solar System if the
sun were to vanish?  Nothing simpler.  Plug in the new values for the
sun's size and mass (\emph{viz}., zero), compute the new orbits, and
compare them to the original ones by using the background simultaneity
and affine structures as referential framework.

The virtues of this method for the representation, interpretation and
evaluation of counterfactuals in the context of physical theory are
legion.  For our purposes, perhaps the greatest virtue is this:
pragmatics plays absolutely no role.  It does not matter what the
investigator's purposes are, or the exact experimental techniques one
envisions as being employed, or anything of the sort.  There is no
need for the \emph{ad hoc} fixing of comparison classes of systems, or
for the \emph{ad hoc} fixing of methods for identifying ``the same
quantity of the same system at the same place and same time, under
otherwise different conditions or in otherwise different states''.
It's all fixed naturally and canonically from the start.\footnote{This
  is not to say there no choices at all to be made in this context.
  One might decide in the Newtonian example, for instance, to consider
  only smooth solutions, or to allow continuous or even distributional
  solutions, in the fixing of the comparison class.  This choice,
  however, does not affect the naturalness and canonicity of the
  procedure.}

Famously in general relativity, there is no fixed, absolute,
background structure one could use as such a referential framework for
the comparison of properties of different solutions to the Einstein
field equation.  Thus, \emph{prima facie}, there is no way to employ
the standard machinery of the initial-value formulation (or the
boundary-value problem) to represent, render meaning to and evaluate
such counterfactuals in general relativity.

\section{The Problem in Detail}
\label{sec:prob-detail}

To begin to come to grips with the problem in more detail, let us
consider the problem of the sun's vanishing translated from the
Newtonian context into that of general relativity: what would
Schwarzschild spacetime look like if its central mass were removed?
Perhaps the immediate, intuitive (and naive) response is to say:
obviously the result will be Minkowski spacetime.  The plethora of
available vacuum solutions, however, shows the choice is not so
simple.

To see the issues involved more clearly, let us try to articulate the
problem in a more explicit, precise and rigorous fashion, using the
idea of taking the limit of a continuously varying family of
spacetimes.  We start with Schwarzschild spacetime of a given mass,
and consider a family of Schwarzschild spacetimes parametrized by mass
continuously shrinking to zero.  It may be initially surprising to
learn that such a limiting family has no unique limit.  (This follows
immediately from the fact that any reasonable topology on a
non-trivial family of spacetimes is non-Hausdorff; see,
\emph{e}.\emph{g}., \citeNP{curiel-meas-topo-prob-cosmo}.)  Working
the example out in a little detail shows clearly what is going
on.\footnote{The following analysis is taken from
  \citeN{geroch-lim-sts}.}  In Schwarzschild coordinates, using the
parameter $\lambda \equiv M^{-1/3}$ (the inverse-third root of the
Schwarzschild mass), the metric takes the form
\begin{equation}
  \label{eq:schwarzschild}
  \left( 1 - \frac{2}{\lambda^3 r} \right) dt^2 - \left( 1 -
    \frac{2}{\lambda^3 r} \right)^{-1} dr^2 - r^2 (d\theta^2 + \sin^2
  \theta \smidge d\phi^2)
\end{equation}
This clearly has no well defined limit as $\lambda \rightarrow 0$.
Now, apply the coordinate transformation 
\[
\tilde{r} \equiv \lambda r, \quad \tilde{t} \equiv \lambda^{-1} t,
\quad \tilde{\rho} \equiv \lambda^{-1} \theta
\]
In these coordinates, the metric takes the form
\[
\left( \lambda^2 - \frac{2}{\tilde{r}} \right) d\tilde{t}^2 - \left(
  \lambda^2 - \frac{2}{\tilde{r}} \right)^{-1} d\tilde{r}^2 -
\tilde{r}^2 (d\tilde{\rho}^2 + \lambda^{-2} \sin^2 (\lambda
\tilde{\rho}) \smidge d\phi^2)
\]
The limit $\lambda \rightarrow 0$ exists and yields
\[
-\frac{2}{\tilde{r}} d\tilde{t}^2 + \frac{\tilde{r}}{2} d\tilde{r}^2 -
\tilde{r}^2 (d\tilde{\rho}^2 + \tilde{\rho}^2 \smidge d\phi^2)
\]
a flat solution discovered by \citeN{kasner-geom-thms-efe}.  If
instead of that coordinate transformation we apply the following to
the original Schwarzschild form~\eqref{eq:schwarzschild},
\[
x \equiv r + \lambda^{-4}, \quad \rho \equiv \lambda^{-4} \theta
\]
then the resulting form also has a well defined limit, which is the
Minkowski metric.  The two limiting processes yield different
spacetimes because it happens behind the scenes that ``the same points
of the underlying manifold get pushed around relative to each other in
different ways''.  Because the coordinate relations of initially
nearby points differ in different coordinate systems, those
differences get magnified in the limit, so that their final metrical
relations differ.  Thus, the limits in the different coordinates yield
different metrics, with no natural or preferred way to say which is
the ``correct'' limit---which, if either, is the correct spacetime
that results from counterfactually removing the sun from the Solar
System.  

The root of the problem lies in the fact that metrical curvature is
\emph{only} in part informed by the distribution of matter: the Weyl
curvature at a point, exactly that part of the curvature encoding
conformal information, such as what counts as a null vector, is
independent of the value of the stress-energy tensor at that
point---the value of the Weyl tensor, point by point, is not
constrained by the presence or absence of matter.  In regions without
matter, moreover, metrical curvature is governed entirely by the Weyl
tensor.  Still, the Weyl tensor $C^a {}_{bcd}$ is subtlely related to
the distribution of matter at neighboring points, when there is such
matter, in a way that can be made precise by using the Bianchi
identity formulated using the so-called Lanczos tensor.\footnote{The
  Lanczos tensor is defined as follows:
  \begin{equation}
    \begin{split}
      J_{abc} &\coloneq \frac{1}{2} \nabla_{[b} R_{a]c} + \frac{1}{6}
      g_{c[a} \nabla_{b]} R \\
      &= 4 \pi \nabla_{[b} T_{a]c} - \frac{1}{12} g_{c[b} \nabla _{a]}
      T
    \end{split}
  \end{equation} 
  The Bianchi identity may then be rewritten
  \[
  \nabla_n C^n {}_{abc} = J_{abc}
  \]
  Thus the value of the Weyl tensor at a point does depend in an
  indirect way on the distribution of matter at nearby points.}  Thus,
in ``removing matter'' from a spacetime region, there can be no
principled way to determine what the ``remaining curvature'' will be.
One may decide to keep the Weyl tensor the same.  But precisely its
relation to stress-energy by way of the Lanczos tensor means that this
is not an unproblematic way to proceed, and is likely even incoherent
or inconsistent.\footnote{It should therefore be clear that these
  sorts of problem arise not only for counterfactuals involving
  changes in the distribution of matter, but also for any involving
  changes in the curvature more generally.  One may, for example, try
  to consider how the behavior of test-particles in a vacuum spacetime
  would change if one were to ``alter a component of the ambient
  gravitational radiation''.}  One may sum all these issues up by
adverting to the fact that there is not a unique vacuum solution to
the Einstein field equation: the form of the Einstein field equation
for a vacuum region remains the same no matter what the field
equations are for matter immediately outside the vacuum region, but
those exterior matter fields do not suffice to fix the solution for
the vacuum region.\footnote{It was, as a matter of fact, exactly this
  issue that first led Einstein to formulate his infamous Hole
  Argument \cite{curiel-exist-st-struct}.}

Now, the problem with this example is, in fact, even more acute than
the discussion so far shows: there are an innumerable number of
possible spacetimes, not all of them flat, with many different
topologies possible, that one can derive by rigorous limiting
processes from Schwarzschild spacetime as the central mass goes to
zero.\footnote{See, \emph{e}.\emph{g}., \citeN{paiva-et-lims-sts} for
  explicit construction of a few different ones.}  The same problems
already discussed arise for all of these, compounded by the great
number and variety of possible topologies the limits may respectively
take.  Let us say, however, that we have, in some way or other,
decided that we are interested in the limit that takes Schwarzschild
spacetime to Minkowski spacetime, and in particular we are interested
in the ways that geodesics near the event horizon will change over the
course of the limit.  The natural topology of the manifold of
Schwarzschild spacetime is $\mathbb{R}^2 \times \mathbb{S}^2$.  The
natural topology of Minkowski spacetimes, however, is $\mathbb{R}^4$.
Thus, in an intuitive sense spacetime points will ``disappear'' in the
limiting process, because one must ``de-compactify two topological
dimensions'' to derive $\mathbb{R}^4$ from $\mathbb{R}^2 \times
\mathbb{S}^2$.  There are many ways to effect such a
de-compactification; all the simplest, such as that based on
Alexandrov compactification, work by the removal of a point or set of
points from the topological manifold.\footnote{See,
  \emph{e}.\emph{g}., \citeN{kelley-gen-topo} for an account of
  methods of compactification, including the Alexandrov type.}  How is
one to identify which spacetime points in the limit manifold
$\mathbb{R}^4$ are those that were ``close to the event horizon'' in
the original Schwarzschild spacetime manifold $\mathbb{R}^2 \times
\mathbb{S}^2$?

The problems discussed here with the representation, interpretation
and evaluation of counterfactuals in general relativity are not
restricted to the case of removing or otherwise altering matter
distributions.  Similarly problematic modalities, requiring the
fixation of a relevant comparison class of spacetimes and the
identification and comparison of structure and properties across
spacetime models, arise all over the place in general relativity.  I
offer a brief sample.
\begin{enumerate}
  \setlength{\itemsep}{0ex}
    \item the proofs of all the Laws of black-hole mechanics
  \cite{wald-qft-cst}
    \item the proof of the No-Hair theorem for black holes
  \cite{heusler-bh-uniq}
    \item proofs that various causality conditions imply each other,
  \emph{e}.\emph{g}., that the absence of closed timelike curves
  implies the absence of closed causal curves
  \cite{hawking-ellis-lrg-scl-struc-st}
    \item proofs of geodesic theorems
  \cite{geroch-jang-motion-gr,ehlers-geroch-eom-small-bods-gr}
    \item well-posedness (in the sense of Hadamard) of the
  initial-value formulation \cite{wald-gr}
    \item formulation of and arguments for the Cosmic Censorship
  Hypothesis \cite{joshi-cosmic-censor}
    \item formulation of and arguments for Penrose's Conformal
  Curvature Hypothesis \cite{penrose-sing-time-asym}
    \item proof of the Topological Censorship Theorem
  \cite{friedman-et-topol-censor}
    \item the possible extendibility of spacetimes past singularities,
  and the possible hole-freeness of a spacetime in general
  \cite{manchak-sings-holes-exts}
    \item taking ``small'' perturbations off a fixed spacetime,
  \emph{e}.\emph{g}., in considering slight inhomogeneities in the
  cosmological FLRW models \cite{szekeres-inhom-cosm-mods}, or in
  treating slightly aspherical gravitational collapse
  \cite{szekeres-quasisph-grav-coll} or more generally distorted black
  holes \cite{geroch-hartle-distorted-bhs}
    \item the genericity or scarcity (or stability or rigidity) of a
  given spacetime property, \emph{e}.\emph{g}., the existence of
  singularities or of large-scale cosmological structure or of a given
  range of values for a constant of nature
  \cite{curiel-meas-topo-prob-cosmo}
\end{enumerate}

\section{Severity of the Problem}
\label{sec:sever-prob}

There is no single algorithm or reasoning procedure to employ for all
such possible cases of counterfactual (and more general modal)
reasoning in general relativity, and certainly no natural or canonical
one.  One must handle such situations on a case-by-case basis, coming
up with a method to fix the comparison class and similarity structure
in the proposed counterfactual situation in some way that respects the
particular constraints of the project the counterfactual reasoning is
to play a part in, while at the same time making possible the
identification and comparison of structure and properties across
spacetimes in the comparison class: the intervention of pragmatics is
inescapable, and will perforce depend on \emph{ad hoc} constructions
and arguments.

The astute reader, however, may now immediately want to respond that
in many interesting accounts of counterfactuals in the literature,
there are formal devices whose sole purpose is to manage the
pragmatics of their interpretation and evaluation.  For example, this
is exactly what the freedom in the selection of ``similarity''
measures in a Lewisian or Stalnakerian account of counterfactuals is
supposed to accommodate.  But the problem is deeper here in general
relativity.  It is not simply that the choice of a similarity measure
is a matter of pragmatics.  It is rather that ``the choice'' in
general is not a well defined set of alternatives \emph{at all}.  Even
once one has decided in some way or other what the appropriate
comparison class of spacetimes ought to be, and what is to count as
the relevant measure of similarity among the spacetimes, still one
simply does not know how, in a general, abstract way, to make sense of
the idea ``different values or forms of the same structure in
different spacetime models'' (or, in Lewisian terms, in different
possible worlds in a family of mutually accessible ones).  Even
putting aside the problems posed by differing topologies in spacetimes
one may want otherwise to judge similar in some important fashion (as
discussed in \S\ref{sec:prob-detail}), the diffeomorphic freedom
inherent in general relativity makes it in general impossible to say
``what is the same point in different spacetimes''.  There is no
principled reason to use $\mathbb{R}^2 \times \mathbb{S}^2$, say, as
the diffeomorphic presentation of Schwarzschild spacetime rather than
$\mathbb{R}^4$ with a line removed (which is diffeomorphic to
$\mathbb{R}^2 \times \mathbb{S}^2$, but not canonically so)---but how
is one to identify in a principled way a point in $\mathbb{R}^2 \times
\mathbb{S}^2$ with one in $\mathbb{R}^4$ with a line removed?  Thus,
not only does general relativity block the standard methods of
managing the pragmatics of counterfactual interpretation and
evaluation, it blocks the prior, even more fundamental step: the
representation of the counterfactual in the given language (in this
case, the mathematical theory of Lorentz manifolds).

Still, one may want to ask, if this is such a problem, how do
physicists handle it in practice?  For they surely do---otherwise no
progress would ever have been made in any of the problems listed at
the end of \S\ref{sec:prob-detail}.  The answer is that, when engaging
in arguments requiring such counterfactual and more generally modal
reasoning, physicists do use \emph{ad hoc} methods for the fixation of
an appropriate comparison class of spacetimes and similarity measure
on it.  More importantly, they use \emph{ad hoc} methods for the
identification and comparison of structure and properties across the
spacetimes in the fixed class, \emph{e}.\emph{g}., by requiring that
the spacetimes in the family have a fixed topology, or have a fixed
diffeomorphic presentation, or satisfy certain symmetries, and then
fixing a coordinate system for the identification of points across
spacetimes.  Though those methods are \emph{ad hoc} in the strong (and
etymologically literal) sense in that they are applicable only to the
specific context in which they are used, and cannot be generalized in
any way to a method applicable in all situations across the theory
(much less generalized in a canonical or natural way), there are often
strong physical justifications for their use.\footnote{Even in many of
  these cases, however, the method used is (figuratively) \emph{ad
    hoc} in the stronger sense that there is no known physically
  significant justification for it.  This happens most often in those
  problems in which the analysis requires mathematical machinery of an
  even more advanced and heavy-duty sort than one needs in workaday
  general relativity.  For instance, I know of no convincing physical
  interpretation for the Sobolev norms required to prove the
  well-posedness of general relativity's initial-value formulation
  \cite{wald-gr}.  Physicists use them because they work
  mathematically, no other reason.  Indeed, I have never even seen a
  physicist pay lip-service to the idea that the Sobolev norms have
  real physical significance.}  This surely suffices for all the
purposes the practical physicist has in his or her theoretical work,
and only a fool or a philosopher would cavil at it.  For we
philosophers, however, who are interested in the understanding and
comprehension of the foundations of the theory as a whole, the problem
of the interpretation and evaluation of counterfactuals in the context
of the theory cannot be considered settled, or even as having a
promising attack made upon it, until reasonable and plausible methods
applicable to the theory as a whole are constructed.

The problem I expose in this paper is severe: many influential
philosophical approaches to many fundamental problems and issues in
the philosophy of science---the meaning of fundamental interpretive
principles, the nature of scientific laws, the stability of
theory-confirmation, the nature of causation, \emph{et al}.--- rely in
ineliminable ways on subjunctive conditionals for their formulation,
analysis and application.  Physicists certainly rely on such
propositions in theoretical and experimental practice all the time,
\emph{e}.\emph{g}., to propose and perform tests of general
relativity.  What reason do we have to believe that we understand what
is happening in such cases in the context of general relativity, much
less to have confidence in any conclusions drawn?  Because the problem
arises solely from the dynamical nature of spacetime geometry in
general relativity, moreover, what I say here is independent of one's
favorite account of counterfactuals---it depends only on the
theoretical resources general relativity provides to model such
situations and pose such propositions, no matter what ancillary tools
or frameworks one uses to interpret and understand them.

I wanted in this paper only to draw attention to this serious problem,
not to propose possible solutions.  I think any decent attempt to do
the latter will require a great deal of involved, technical work,
including detailed examination of many non-trivial examples.  I
sincerely hope someone takes up the challenge.  I am not entirely
without hope for progress to be made here.  Indeed, it seems likely
that progress on this problem could suggest new avenues of attack on
the traditional problem of understanding counterfactuals in general,
and could suggest elaborations of and improvements to already existing
accounts.  In any event, beyond the possible benefits that progress on
this problem may have for the general problem of counterfactuals, the
mettle of philosophy and the needs of physics demand we understand
what is going on here.

\addcontentsline{toc}{section}{\hspace*{-1.3em}\numberline{}References}

\bibliographystyle{chicago}

\end{document}